\documentclass[showpacs,showkeys,preprintnumbers,amsmath,amssymb]{revtex4}

\usepackage{graphicx}
\usepackage{dcolumn}
\usepackage{bm}

\begin{document}

\title{TIME DURATION OF THE PARAMETRIC X-RAY RADIATION}

\author{S. V. Anishchenko}
\email{sanishchenko@mail.ru}
\affiliation{Research Institute for Nuclear Problems\\
Bobruiskaya str., 11, 220050, Minsk, Belarus.}%

\author{V. G. Baryshevsky}
 \email{bar@inp.minsk.by}
 \affiliation{Research Institute for Nuclear Problems\\
Bobruiskaya str., 11, 220050, Minsk, Belarus.}%

\author{A.A. Gurinovich}
 \email{gur@inp.minsk.by}
 \affiliation{Research Institute for Nuclear Problems\\
Bobruiskaya str., 11, 220050, Minsk, Belarus.}%


\begin{abstract}
Time evolution of the parametric X-Ray radiation, produced by a relativistic charged particle passing through a crystal, is studied. The most attention is given to the cases when the radiation lasts much longer ($t_{PXR}\sim0.1$~ns) than the the time
$t_p$ of  the particle flight through the crystal ($t_p\sim1$~ps). It is shown that such long duration of the radiation makes possible the detailed experimental investigation of the complicated time structure of the parametric X-ray pulses, generated by electron bunches, which are available with modern acceleration facilities.
\end{abstract}




\maketitle

\section*{Introduction}
The investigation of stop-actions in the femtosecond time scale is one of the most important part of research programs on modern accelerators, used for X-ray lansing (FLASH \cite{flash}, XFEL \cite{xfel}, LCLS \cite{lcls}, SPring-8 \cite{spring8}, SwissFEL~\cite{swissfel}). Being able to produce femtosecond electron bunches the accelerators can provide researchers with important information about X-ray radiation generated by charged relativistic particles moving in crystals \cite{VG1989,VG1999,VGAA2011}.

It is well known the refraction of the electromagnetic field associated with the electron passing through matter with a uniform velocity originates the Vavilov-Cherenkov and transition radiations \cite{Landau7}. The Cherenkov emission of photons by a charged particle occurs whenever the index of refraction $n(\omega)>1$ ($\omega$ is the photon frequency). For X-ray frequencies being higher than typical atom's ones the refractive index has a universal form \cite{Landau7}
\begin{equation}
n(\omega)=1-\frac{\omega_L^2}{2\omega^2}
\end{equation}
where $\omega_L$ is the Langmuir frequency.
As a result, $n(\omega)>1$ and the Vavilov-Cherenkov effect is seemed to be absent in the X-ray range.

But as it was first shown in \cite{VG1971,VGID1971} the diffraction of photons emitted by a relativistic charged particle in crystals leads to an abrupt change in the refractive index of X-ray quanta and, as a result, gives rise to the the spontaneous and induced quasi-cherenkov radiation called the parametric X-ray radiation (PXR). It also leads to a considerable change in spectral-angular characteristics of transition and bremsstrahlung radiation at both small and large angles to the direction of particle motion \cite{VG1971,VGOM2003}.

In accordance with \cite{VG1971,VGID1971} the diffraction of virtual photons in a crystal can be described by a set of refractive indices $n_\mu(\omega,\vec k)$, some of which may appear to be greater than unity. Here $\vec k$ is the refraction index of a crystal for the X-rays with the photon wave vector $\vec k$. Particularly, in the case of the two wave diffraction, the refractive indices are $n_1(\omega,\vec k)>1$ and $n_2(\omega,\vec k)<1$, and accordingly, the two types of waves propagate in a crystal: a fast wave ($n_2<1$) and a slow one ($n_1>1$). For a slow wave the Cherenkov condition can be fulfilled, but not for a fast wave. The latter wave is emitted at the vacuum-crystal boundary (or due to bremsstrahlung radiation at multiple scattering).

At present there are lots of theoretical and experimental works devoted to the PXR being first registred experimentaly in 1985 \cite{BFU}.  Moreover the investigation of the tunable sources of radiation based on the parametric radiation are in progress.

It should be noted, however, that until now,
theoretical and experimental analysis of radiation produced by a
relativistic particle passing through a crystal has focused on
spectral-angular characteristics of radiation. Nevertheless, it
was shown in \cite{VG1989,VG1999,VGAA2011} that because of diffraction,
emitted photons have the group velocity
$v^{p}_{gr}$ appreciably smaller than the velocity $v$
of a relativistic particle.  As a result, the situation is
possible in which radiation from the crystal still continues after
the particle has passed through it \cite{VG1989,VG1999,VGAA2011}. This
enables studying  time evolution of the process of photon
radiation  produced during particle transmission through
a natural or artificial (photonic) crystal, or during the particle flight along
the surface of such crystals \cite{VG1989,VG1999,VGAA2011}. In the present paper the formulas
are derived, which describe the time evolution of the parametric X-ray radiation
produced by a relativistic particle moving in a crystal. It is
shown that the conditions are realizable under which the PXR lasts considerably longer than the time
$t_p$ of  the particle flight through the crystal.

Time resolution of modern X-ray detectors used in FEL technology lies in the picosecond range \cite{Det1,Det2}. Consequently time dependence of the parametric radiation can be investigated experimentaly.

The paper is organized as follows. In Sec. 1 formula for the group velocity of photons in crystals is derived in the two wave approximation. It is shown that the group velocity can be much smaller than the speed of light in the vacuum. Long time reflection of the X-ray pulses from crystals demonstrates this (Sec. 2). Equations, describing the time evolution of the PXR, are derived in sec. 3. The dependence of X-ray pulses on geometrical parameters are analysed.

\section{Group velocity of the electromagnetic waves in crystals}
Let us consider the pulse of electromagnetic radiation passing through the
medium with the index of refraction $n(\omega )$. The  group
velocity of the wave packet is as follows:

\begin{equation}
v_{gr}=\left( \frac{\partial \omega n(\omega )}{c\partial \omega }\right)
^{-1}=\frac{c}{n(\omega )+\omega \frac{\partial n(\omega )}{\partial \omega }%
},
\label{v1}
\end{equation}
where $c$  is the speed of light, $\omega $  is the quantum frequency.

In the X-ray range ( $\sim $tens of keV) the index of refraction has the
universal form $n(\omega )=1-\frac{\omega _{L}^{2}}{2\omega ^{2}}$ , $\
\omega _{L}$ is the Langmuir frequency. Additionally, $n-1\simeq 10^{-6}\ll 1
$. Substituting \ $n(\omega )$ into (\ref{v1}), one can obtain that $v_{gr}\simeq
c\left( 1-\frac{\omega _{L}^{2}}{\omega ^{2}}\right) $. It is clear that
the group velocity is close to the speed of light. Therefore the time delay of
the wave packet  in a medium is much shorter than the time needed for
passing the path equal to the target thickness $L$ in the vacuum.

\begin{equation}
\Delta T=\frac{L}{v_{gr}}-\frac{L}{c}\simeq \frac{L}{c}\frac{\omega _{L}^{2}%
}{\omega ^{2}}\ll \frac{L}{c}.
\label{v2}
\end{equation}

To consider the pulse diffraction in a crystal, one should solve Maxwell's
equations that describe a pulse passing through a crystal. Maxwell's equations
are linear, therefore it is convenient to use the Fourier transform in time and
to rewrite these equations as functions of frequency:
\begin{equation}
\left[ -curl~curl~\vec{E}_{\vec{k}}(\vec{r},\omega )+\frac{\omega ^{2}}{c^{2}%
}\vec{E}_{\vec{k}}(\vec{r},\omega )\right] _{i}+\chi _{ij}(\vec{r},\omega
)~E_{\vec{k},j}(\vec{r},\omega )=0,
\label{v3}
\end{equation}
where $\chi _{ij}(\vec{r},\omega )$  is the spatially periodic
tensor of susceptibility; $i,j=1,2,3$ repeated  indices imply
summation.

Making the Fourier transformation of these equations in coordinate
variables, one can derive a set of equations associating the
incident and diffracted waves. When two strong waves are excited
under diffraction (the so-called two-beam diffraction case), the
following set of equations for determining the wave amplitudes can
be obtained:
\begin{equation}
\begin{array}{c}
\left( \frac{k^{2}}{\omega ^{2}}-1-\chi _{0}\right) \vec{E}_{\vec{k}%
}^{s}-c_{s}\chi _{-\vec{\tau}}\vec{E}_{\vec{k}_{\tau }}^{s}=0 \\
\\
\left( \frac{k_{\tau }^{2}}{\omega ^{2}}-1-\chi _{0}\right) \vec{E}_{\vec{k}%
_{\tau }}^{s}-c_{s}\chi _{\vec{\tau}}\vec{E}_{\vec{k}}^{s}=0
\label{v4}
\end{array}
\end{equation}
Here $\vec{k}$ is the wave vector of the incident wave, $\vec{k}_{\vec{\tau}%
}=\vec{k}+\vec{\tau}$, $\vec{\tau}$ is the reciprocal lattice vector; $\chi
_{0},\chi _{\vec{\tau}}$ are the Fourier components of the crystal
susceptibility:
\begin{equation}
\chi (\vec{r})=\sum_{\vec{\tau}}\chi _{\vec{\tau}}\exp (i\vec{\tau}\vec{r})
\label{v5}
\end{equation}
$C_{s}=\vec{e}^{~s}\vec{e}_{\vec{\tau}}^{~s}$, $\vec{e}^{~s}(\vec{e}_{\vec{%
\tau}}^{~s})$ are the unit polarization vectors of the incident and
diffracted waves, respectively.

The solvability condition for the linear system (\ref{v4}) leads to a dispersion
equation that determines the possible wave vectors $\vec{k}$ in a crystal \cite{Pinsker,Chang}.
It is convenient to present these wave vectors as:

\begin{equation}
\vec{k}_{\mu s}=\vec{k}+\texttt{\ae }_{\mu s}\vec{N},~\ae _{\mu s}=\frac{%
\omega }{c\gamma _{0}}~\varepsilon _{\mu s},
\label{v6}
\end{equation}

where $\mu =1,2$; $\vec{N}$ is the unit vector of a normal to the entrance surface
of the crystal, which is directed into the crystal,
\begin{equation}
\varepsilon _{s}^{(1,2)}=\frac{1}{4}[(1+\beta_1 )\chi _{0}-\beta_1 \alpha _{B}%
]\pm \frac{1}{4}\left\{ [(1+\beta_1 )\chi _{0}-\beta_1 \alpha _{B}-2\chi
_{0}]^{2}+4\beta_1 C_{S}^{2}\chi _{\vec{\tau}}\chi _{-\vec{\tau}}\right\}
^{1/2},
\label{v7}
\end{equation}
$\alpha _{B}=(2\vec{k}\vec{\tau}+\tau ^{2})k^{-2}$ is the off-Bragg
parameter ($\alpha _{B}=0$ when the Bragg condition of diffraction is
exactly fulfilled),
$
\gamma _{0}=\vec{n}_{\gamma }\cdot \vec{N},~~~\vec{n}_{\gamma }=\frac{\vec{k}%
}{k},~~~\beta =\frac{\gamma _{0}}{\gamma _{1}},~~~\gamma _{1}=\vec{n}%
_{\gamma \tau }\cdot \vec{N},~~~\vec{n}_{\gamma \tau }=\frac{\vec{k}+\vec{%
\tau}}{|\vec{k}+\vec{\tau}|}$, $C_s=1$ ($\sigma$-polarization), $C_s=\cos(2\theta_B)$ ($\pi$-polarization, $\theta_B$ --- Bragg angle).

The general solution of equations (\ref{v3}), (\ref{v4}) inside a crystal is:
\begin{equation}
\vec{E}_{\vec{k}}^{s}(\vec{r})=\sum_{\mu =1}^{2}\left[ \vec{e}^{~s}A_{\mu
}\exp (i\vec{k}_{\mu s}\vec{r})+\vec{e}_{\tau }^{~s}A_{\tau \mu }\exp (i\vec{%
k}_{\mu s\tau }\vec{r})\right]
\label{v8}
\end{equation}

Associating these solutions with the solutions of Maxwell's
equation for the
vacuum area one can find the explicit expression for $\vec{E}_{\vec{k}}^{s}(%
\vec{r})$ throughout the space. It is possible to discriminate several types
of diffraction geometries, namely, the Laue and the Bragg schemes, which  are
most well-known \cite{Pinsker}.

In the case of two-wave dynamical diffraction, the crystal can be described \ by
two effective  indices of refraction

\[
n_{s}^{(1,2)}=1+\varepsilon _{s}^{(1,2)},
\]

\begin{equation}
\varepsilon _{s}^{(1,2)}=\frac{1}{4}\left\{ \chi _{{\small 0}}(1+\beta_1
)-\beta_1 \alpha_B \pm \sqrt{(\chi _{{\small 0}}(1-\beta_1 )+\beta_1 \alpha_B
)^{2}+4\beta_1 C_{s}\chi _{{\small \tau }}\chi _{{\small -\tau }}}\right\} .
\label{v9}
\end{equation}

The diffraction is significant in the narrow range near the Bragg frequency,
therefore $\chi _{0}$ and $\chi _{\tau }$ can be considered as constants
and\ the dependence on $\omega $ should be taken into account for \ $\alpha_B=%
\frac{\overrightarrow{\tau }(\overrightarrow{\tau }+2%
\overrightarrow{k})}{k^{2}}=-\frac{\tau^{2}}{k_{B}^{3}c}(\omega
-\omega _{B})$, where $k=\frac{\omega }{c}$; $\overrightarrow{\tau }$\
is the reciprocal lattice vector which characterizes the set of planes where
the diffraction occurs; Bragg frequency is determined by the condition $%
\alpha_B =0$, i. e. $\tau^2+2\overrightarrow k_B\overrightarrow\tau=0$ and $\omega_B=-\frac{\tau^{2}c}{2\overrightarrow n_\gamma\overrightarrow\tau}$.

From (\ref{v1}), (\ref{v8})\ one can obtain

\begin{equation}
\begin{split}
&\frac{v_{gr}^{(1,2)s}}{c}=\frac{1}{n_s^{(1,2)}+\omega\frac{d\alpha_B}{d\omega}\frac{\partial n_s^{(1,2)}}{\partial\alpha_B}+\omega\frac{d\chi_0}{d\omega}\frac{\partial n_s^{(1,2)}}{\partial\chi}+\omega\frac{d(\chi_\tau\chi_{-\tau})}{d\omega}\frac{\partial n_s^{(1,2)}}{\partial(\chi_\tau\chi_{-\tau})}}=\\
&\frac{1}{n_s^{(1,2)}+\frac{\omega}{4}\bigg(\frac{d\chi}{d\omega}(1+\beta_1)-\beta_1\frac{d\alpha_B}{d\omega}\pm\frac{\big(\chi_0(1-\beta_1)+\beta_1\alpha_B\big)\big(\frac{d\chi_0}{d\omega}(1-\beta_1)+\frac{d\alpha_B}{d\omega}\beta_1\big)+2\beta_1 C_s^2\frac{d(\chi_\tau\chi_{-\tau})}{d\omega}}{\sqrt{\big(\chi_0(1-\beta_1)+\beta_1\alpha_B\big)^2+4\beta C_s^2\chi_\tau\chi_{-\tau}}}\bigg)}\\
\end{split}
\label{v10}
\end{equation}

From \eqref{v10} it follows, that the complicated character of the wave field in a crystal leads to both the negative and the positive values of the group velocity. Moreover  when $\beta_1$ is negative the radicand in (\ref{v10}) approaches zero
(Bragg reflection threshold) and $v_{gr}\rightarrow 0$\ . The performed analysis allows one to
conclude that the center of the X-ray pulse in a crystal can \
undergo a significant delay $\Delta T\gg \frac{L}{c}$ available
for experimental investigation. Thus, when $\beta_1=10^{2}$,
$L=0.3$ mm and $L/c\simeq 1$~ps, the delay \ time can
be estimated as $\Delta T\approx\frac{L}{v_{gr}}\simeq 0.1$~ns.

%


\begin{figure}
 \begin{center}
 \begin{tabular}{c}
      \resizebox{80mm}{!}{\includegraphics{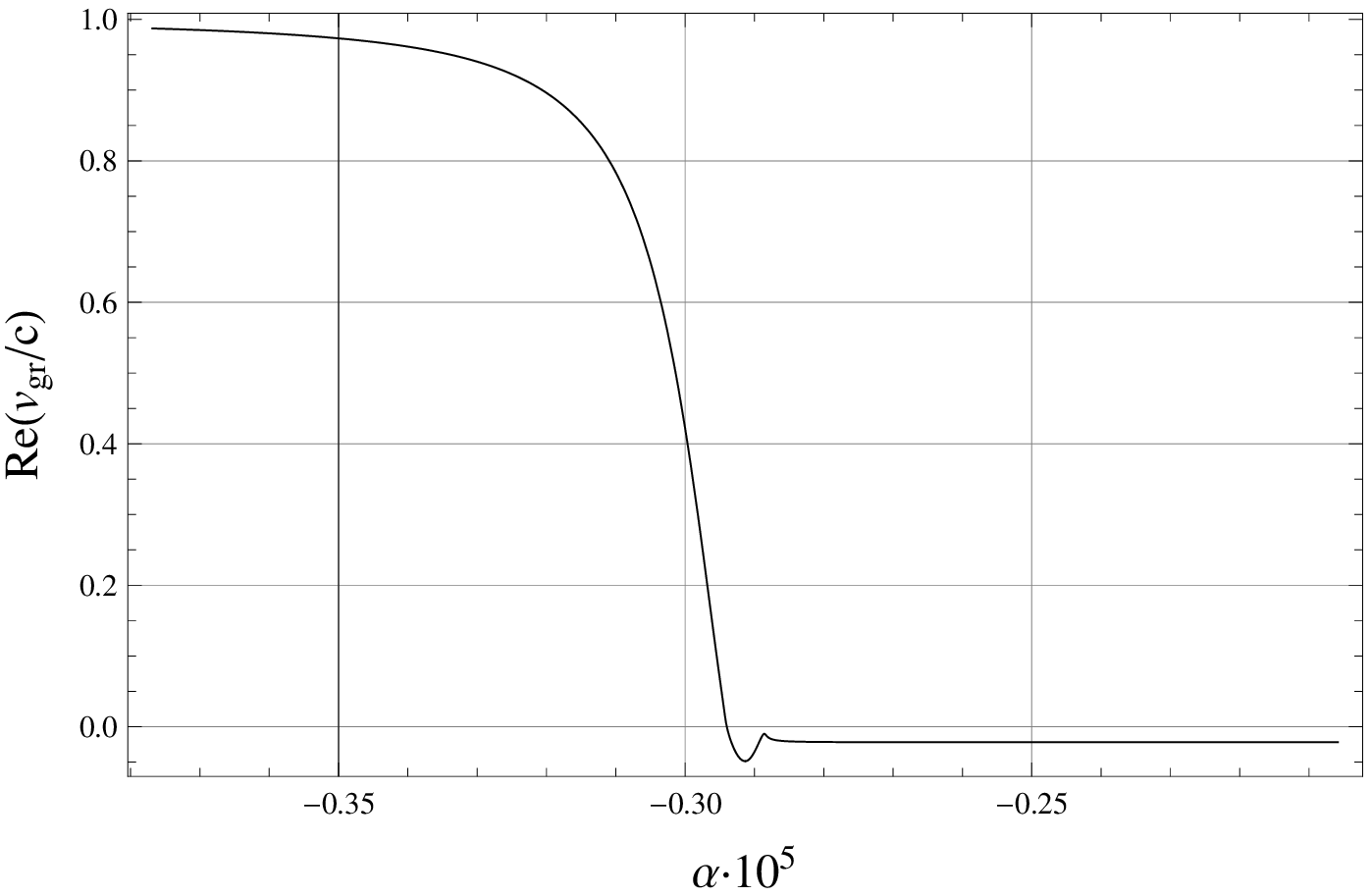}}\\
\end{tabular}

\small{Fig. 1. Group velocity of the slow wave for the Bragg diffraction case in LiF crystal ($\beta_1=-45$).}

 \end{center}
\end{figure}

\section{Wave-packet reflection}

Let us study now the time dependence of intensity of the wave-packet \ after
passing through a crystal more datail. Assuming that $B(\omega )$ is the reflection or
transmission amplitude coefficients of a crystal, one can obtain the
following expression for the pulse form

\begin{equation}
E(t)=\frac{1}{2\pi }\int B(\omega )E_{0}(\omega )e^{-i\omega t}d\omega =\int
B(t-t^{\prime })E_{0}(t^{\prime })dt^{\prime }.
\label{v11}
\end{equation}
where $E_{0}(\omega )$ is the amplitude of the electromagnetic wave incident
on a crystal.

In accordance with the general theory, for the Bragg geometry, the
amplitude of the diffraction-reflected wave for the crystal width
 much greater than the absorbtion length can be written as
 \cite{Pinsker}:

\begin{eqnarray}
& &B_{s}(\omega )=\\
& &-\frac{1}{2\chi _{\tau }}\left\{ \chi _{{\small 0}}(1+\left|
\beta_1 \right| )-\left| \beta_1 \right| \alpha_B -\sqrt{(\chi _{{\small 0}%
}(1-\left| \beta_1 \right| )-\left| \beta_1 \right| \alpha_B )^{2}-4\left| \beta_1
\right| C_{s}\chi _{{\small \tau }}\chi _{{\small -\tau }}}\right\}\nonumber
\label{v12}
\end{eqnarray}

In the absence of resonance scattering, the parameters $\chi _{0}$
and $\chi _{\pm \tau }$ can be considered as \ constants and frequency
dependence is defined by the term $\alpha_B=-\frac{\tau^{2}}{k_{B}^{3}c}(\omega
-\omega _{B})$.

So, $B_{s}(t)$\ \ can be found from

\begin{eqnarray}
& &B_{s}(t)=-\frac{1}{4\pi \chi _{\tau }}\\
& &\times \int \left\{ \chi _{{\small 0}%
}(1+\left| \beta_1 \right| )-\left| \beta_1 \right| \alpha_B -\sqrt{(\chi _{%
{\small 0}}(1-\left| \beta_1 \right| )-\left| \beta \right| \alpha_B
)^{2}-4\left| \beta_1 \right| C_{s}\chi _{{\small \tau }}\chi _{{\small -\tau }%
}}\right\} e^{-i\omega t}d\omega .\nonumber
\label{v13}
\end{eqnarray}

The Fourier transform of the first term results in $\delta (t)$ and we can
neglect it because the delay is described by the second term. The second
term can be calculated by the methods of the theory of \ function of complex
argument:

\begin{equation}
B_{s}(t)=-\frac{i}{4\chi _{\tau }}\left| \beta_1 \right|\frac{\tau^{2}}{k_{B}^{2}\omega_B}\frac{J_{1}(a_{s}t)}{t}e^{-i(\omega _{B}+\Delta
\omega _{B})t}\theta (t),
\label{v14}
\end{equation}

or

\begin{equation}
B_{s}(t)=-\frac{i\sqrt{\left| \beta_1 \right| }}{2}\frac{J_{1}(a_{s}t)}{a_{s}t}%
e^{-i(\omega _{B}+\Delta \omega _{B})t}\theta (t),
\label{v15}
\end{equation}

where

\[
a_{s}=\frac{2\sqrt{C_{s}\chi _{\tau }\chi _{-\tau }}\omega _{B}}{\sqrt{%
\left| \beta_1 \right| }\frac{\tau^{2}}{k_{B}^{2}}},\Delta \omega
_{B}=-\frac{\chi _{{\small 0}}(1+\left| \beta_1 \right| )\omega_Bk_B^2}{%
\left| \beta_1 \right|\tau^2}.
\]

Since $\chi _{0}$ and $\chi _{\tau }$ are complex, both $a_{s}$
and $\Delta \omega _{B}$ have real and imaginary parts. According
to (\ref{v12})--(\ref{v14}), in the case of Bragg reflection of a
short pulse (the pulse frequency bandwidth $\gg $ frequency
bandwidth of the total reflection range)  both the instantly
reflected pulse and the pulse with amplitude undergoing damped
beatings appear. Beatings period increases with $\left| \beta_1
\right| $ grows and $\chi _{\tau }$\ decrease. Pulse intensity can
be written as

\begin{equation}
I_{s}(t)\sim \left| B_{s}(t)\right| ^{2}=\frac{\left| \beta_1 \right| }{2}%
\left| \frac{J_{1}(a_{s}t)}{at}\right| ^{2}e^{-2\texttt{Im}\Delta \omega
_{B}t}\theta (t).
\label{v16}
\end{equation}

It is evident that the reflected pulse intensity depends on the orientation
of photon polarization vector $\vec{e}_{s}$ and undergoes the
damping oscillations on time. \qquad \qquad\ \

Let us evaluate the effect. Characteristic values are $\texttt{Im}\Delta
\omega _{B}\sim \texttt{Im}\chi _{0}\omega _{B}$ and $\texttt{Im}a\sim \frac{%
\texttt{Im}\chi _{\tau }\omega _{B}}{\sqrt{\beta_1 }}.$ For 10 keV for the
crystal of Si $\ \texttt{Im}\chi _{0}=1,6\cdot 10^{-7}$ , \ for LiH $\ \texttt{Im%
}\chi _{0}=7,6\cdot 10^{-11},\texttt{Im}\chi _{\tau }=7\cdot 10^{-11}$, \ for
LiF $\ \texttt{Im}\chi _{0}\sim 10^{-8}.$ Consequently, the characteristic
time $\tau $\ for the exponent decay in (\ref{v15}) can be estimated as follows ($%
\omega _{B}=10^{19}$):

for Si the characteristic time $\tau \sim 10^{-12}$ sec, for LiF
the characteristic time $\tau \sim 10^{-10}$ sec, for LiH the
characteristic time $\tau \sim 10^{-9}$ sec!!

The reflected pulse also undergoes oscillations, the period of
which increases with growing $\beta_1$ and
decreasing $\texttt{Re}\chi _{\tau }. $ This period can be
estimated for $\beta =10^{2}$ and $\texttt{Re}\chi _{\tau }\sim
10^{-6}$ as $T \sim 10^{-12}$ sec (for Si, LiH, LiF).

When the resolving time of the detecting equipment is greater than
the oscillation period, the expression (\ref{v15}) should be
averaged over the
period of oscillations and the delay law (\ref{v15}) has the
power function form:

\[
{\large I}_{s}{\large (t)\,\sim \,t}^{-3}{\large}e^{-2\texttt{Im}\Delta\omega_Bt}.
\]

In  the case of multi-wave diffraction, the time delay for the photon exit from the crystal will be even more.

\section{Parametric X-Ray radiation}

\begin{figure}
 \begin{center}
 \begin{tabular}{c}
      \resizebox{80mm}{!}{\includegraphics{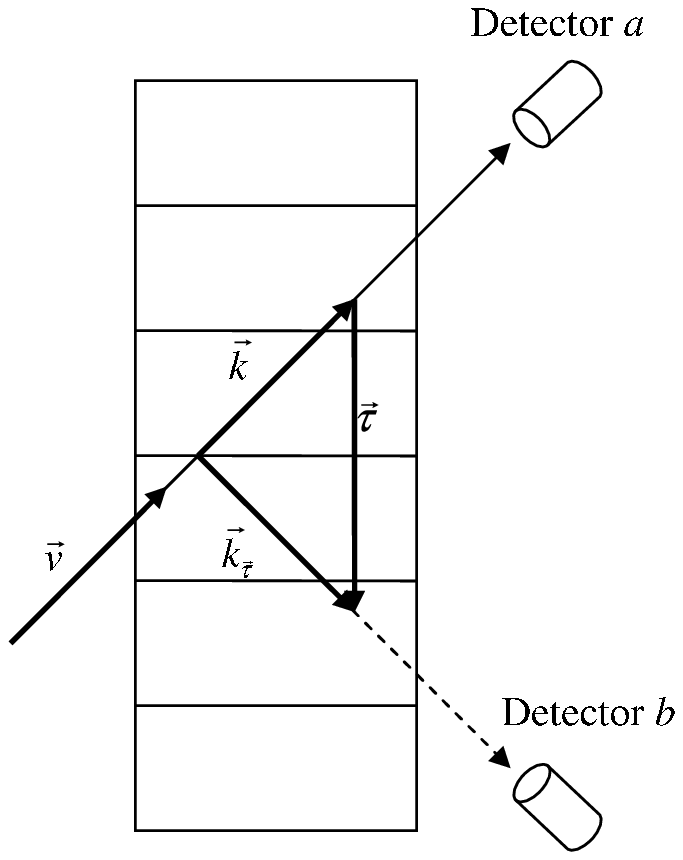}}\\
\end{tabular}

\small{Fig. 2. Diffraction in the Laue geometery: forward (\textit{a}) and diffracted (\textit{b}) maxima.}

 \end{center}
\end{figure}

\begin{figure}
 \begin{center}
 \begin{tabular}{c}
      \resizebox{80mm}{!}{\includegraphics{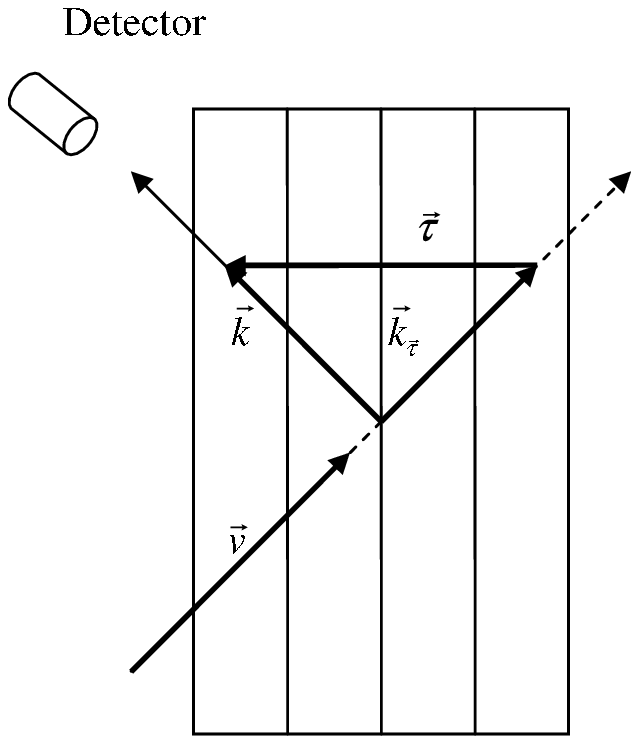}} \resizebox{80mm}{!}{\includegraphics{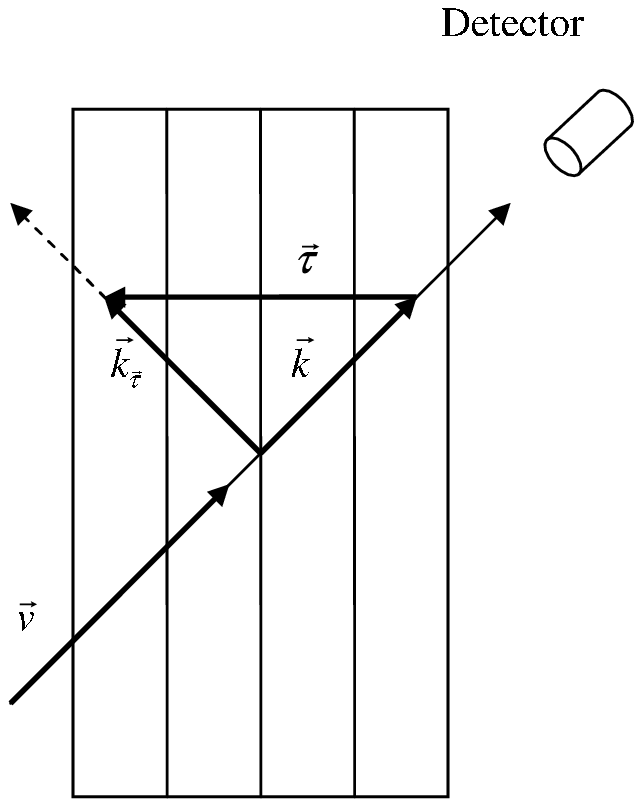}}\\
\end{tabular}

\small{Fig. 3. Schemes of diffraction in the Bragg geometeries: diffracted maximum (left),  forward maximum (right).}

 \end{center}
\end{figure}

So the time $t_{ph}=\frac{L}{v_{gr}}$ that photons spend in a crystal can be longer than the flight time $t_p=\frac{L}{v}$ of a
 relativistic particle in a crystal. Hence, the emission of diffraction-related PXR (diffracted radiation of an oscillator, surface parametric radiation and others) produced by a
 relativistic particle will continue after the particle has left the crystal.
 Under diffraction conditions,
 the crystal acts as a high-quality resonator.

The intensity $I(t)$ of radiation produced by a particle which has passed through a crystal can be found with known intensity of the electric field $\vec{E}(\vec{r},t))$  (magnetic field $\vec{H}(\vec{r},t)$) of the electromagnetic wave, which is produced by this particle \cite{Landau7},
\begin{equation}
\label{1.1}
I(t)=\frac{c}{4\pi}|\vec{E}(\vec{r},t)|^2r^2,
\end{equation}
where $r$ is the distance from the crystal, which is assumed to be larger than the crystal size.

The field $\vec{E}(\vec{r},t)$ can be presented as an expansion in a Fourier series
\begin{equation}
\label{1.2}
\vec{E}(\vec{r},t)=\frac{1}{2\pi}\int\vec{E}(\vec{r},\omega)e^{-i\omega t}d\omega.
\end{equation}
At a long distance from the crystal, the Fourier component $\vec E(\vec r,\omega)$ can be written as follows \cite{VGChanneling,VG1997,BFU}:
\begin{equation}
\label{1.3}
E_i(\vec{r},\omega)=\frac{e^{ikr}}{r}\frac{i\omega}{c^2}\sum_{s}e_i^s
\int\vec{E}_{\vec{k}}^{(-)s^*}(\vec{r}^{\prime}\omega)\vec{j}(\vec{r}^{\prime},\omega)  d^3 r^{\prime}.
\end{equation}
where $i=1,2,3$ (and correspond(s) to the coordinate axes $x$, $y$, $z$), $e^s_i$ is the $i$-component  of the wave polarization vector $\vec{e}^s$; $s=1,2$; $\vec{E}^{(-)s}_{\vec{k}}$ is the solution of  Maxwell's equations,
\begin{equation}
\label{1.4}
\vec{j}(\vec{r}, \omega)=\int \vec{j}(\vec{r}, t)e^{i\omega t}dt
\end{equation}
$\vec{j}(\vec{r}, \omega)= Q\vec{v}(t)\delta(\vec{r}-\vec{r}(t))$ is the is the current density of the particle with charge $Q$, $\vec{r}(t)$ is the particle coordinate at time $t$.

The explicit form of the expressions $\vec{E}^{(-)s}_{\vec k}$ describing
diffraction of the electromagnetic wave in a crystal in the Laue
and Bragg cases is given in \cite{VGChanneling,VG1997}.
It should be noted there is the close connection between $\vec E^{(-)s}_{\vec k}$ and $\vec E^{(+)s}_{\vec k}$,  describing scattering of a plane wave with a wave vector $\vec{k}=k\frac{\vec{r}}{r}$ on a target and the asymptotic of a diverging spherical wave: $\vec E^{(-)s}_{\vec k}=\vec E^{(+)s*}_{-\vec k}$ \cite{VGChanneling,VG1997,BFU}. This circumstance permits us to use the results obtained in diffraction theory to find explicit expressions for $\vec E^{(-)s}_{\vec k}$.

Now let us look at the expression for the amplitude $A(\omega)$ of the emitted wave more attentively:
\begin{equation}
\label{1.5}
A_{\vec{k}}^s(\omega)=\frac{i\omega}{c^2}\int \vec{E}_{\vec{k}}^{(-)s^*}(\vec{r}^{\prime},\omega)\vec{j}(\vec{r}^{\prime},\omega)d^3 r^{\prime}.
\end{equation}
Using (\ref{1.4}), (\ref{1.5}) can be recast as follows
\begin{eqnarray}
\label{1.61}
A_{\vec{k}}^s(\omega)&=&\frac{i\omega}{c^2}\int \vec{E}_{\vec{k}}^{(-)s^*}(\vec{r}^{\prime},\omega)Q\vec{v}(t)\delta(\vec{r}^{\prime}-\vec{r}(t))e^{i\omega t}dt d^3 r^{\prime}\nonumber\\
&=&\frac{i\omega Q}{c^2}\int\vec{E}_{\vec{k}}^{(-)s^*}(\vec{r}(t),\omega)\vec{v}(t)e^{i\omega t} dt
\end{eqnarray}

According to (\ref{1.61}), the radiation amplitude is
determined by the field $\vec{E}_{\vec{k}}^{(-)s}$ taken at point
$\vec{r}(t)$ of particle location at time $t$ and integrated over
the time of particle motion.

From (\ref{1.2}), (\ref{1.3}), and (\ref{1.5}) it follows that the expression for the electromagnetic wave emitted by the particle passing through the crystal (natural or photonic) can be presented in a form:
\begin{equation}
\label{1.7}
\vec{E}(\vec{r},t)=\frac{1}{2\pi r}\sum_s \vec e^s\int A_{\vec{k}}^s(\omega)e^{-i\omega(t-\frac{r}{c})}d\omega,
\end{equation}
i.e., $\vec{E}_i(\vec{r},t)=\frac{1}{r}\sum\limits_s e_i^s A^s_{\vec{k}}(t-\frac{r}{c})$.

Consequently the expression for the X-ray intensity can be written as
\begin{equation}
I(\vec r,t)=\frac{c}{4\pi}|\sum\limits_s \vec e^s A^s_{\vec{k}}(t-\frac{r}{c})|^2.
\end{equation}

If we are interested in the intensity of emitted photons with definite polarization then
\begin{equation}
I_s(\vec r,t)=\frac{c}{4\pi}|\vec e^s A^s_{\vec{k}}(t-\frac{r}{c})|^2.
\end{equation}

From (\eqref{1.7}) follows that the time dependence of the form of the pulse $I(\vec{r},t)(\vec{E}(\vec{r},t))$ of radiation generated by a particle passing through the crystal is determined by the dependence of the radiation amplitude $A_{\vec{k}}^s(\omega)$ on frequency. According to \eqref{1.61} to find explicite expressions for the radiation amplitudes $A^s_{\vec k}(\omega)$ it is neccesary to know wave functions $\vec E^{(-)s}_{\vec k}$, which have been obtained in \cite{VGChanneling,VG1997,BFU}.

\subsection{Laue case}
Let us consider Laue case (fig. 2). In this case electromagnetic waves emitted by a particle in the forward and diffracted directions leave the crystal through the same surface. By matching the solutions of Maxwell's equations on the crystal surfaces one can obtain the following expression for the amplitude in Laue  case for forward directions \cite{VG1997}
\begin{eqnarray}
\label{1.8}
A_{\vec{k}}^s(\omega)&=&\frac{Q\omega}{c^2}(\vec{e}^s \vec{v})\sum_{\mu=1,2}\xi^0_{\mu s}e^{i\frac{\omega}{\gamma_0}\varepsilon_{\mu s}L}\nonumber\\
&\times&\left[\frac{1}{\omega-\vec{k}\vec{v}}-\frac{1}{\omega-(\vec{k}+\kappa_{\mu s}\vec{N})\vec{v}}\right]\left[e^{i(\omega-(\vec{k}+\kappa_{\mu s}\vec{N})\vec{u})\frac{L}{c\gamma_0}} -1\right]
\end{eqnarray}
and for diffracted directions
\begin{eqnarray}
A_{\vec{k_\tau}}^s(\omega)&=&\frac{Q\omega}{c^2}(\vec{e}^{s\tau} \vec{v})\sum_{\mu=1,2}\xi^\tau_{\mu s}e^{i\frac{\omega}{\gamma_\tau}\varepsilon_{\mu s}L}\nonumber\\
&\times&\left[\frac{1}{\omega-\vec{k_\tau}\vec{v}}-\frac{1}{\omega-(\vec{k_\tau}+\kappa_{\mu s}\vec{N})\vec{v}}\right]\left[e^{i(\omega-(\vec{k_\tau}+\kappa_{\mu s}\vec{N})\vec{u})\frac{L}{c\gamma_0}} -1\right],
\end{eqnarray}

where
\begin{equation}
\begin{split}
 &\overrightarrow k_\tau=\overrightarrow k+\overrightarrow \tau,\\
 &\xi_{1(2)s}^0=\mp\frac{2\epsilon_{2(1)s}-\chi_0}{2(\epsilon_{2s}-\epsilon_{1s})},\\
 &\xi_{1(2)s}^{\vec\tau}=\mp\frac{C_s\chi_{-\vec\tau}}{2(\epsilon_{2s}-\epsilon_{1s})}.\\
\end{split}
\end{equation}

\subsection{Bragg case}

\begin{figure}
 \begin{center}
 \begin{tabular}{c}
      \resizebox{80mm}{!}{\includegraphics{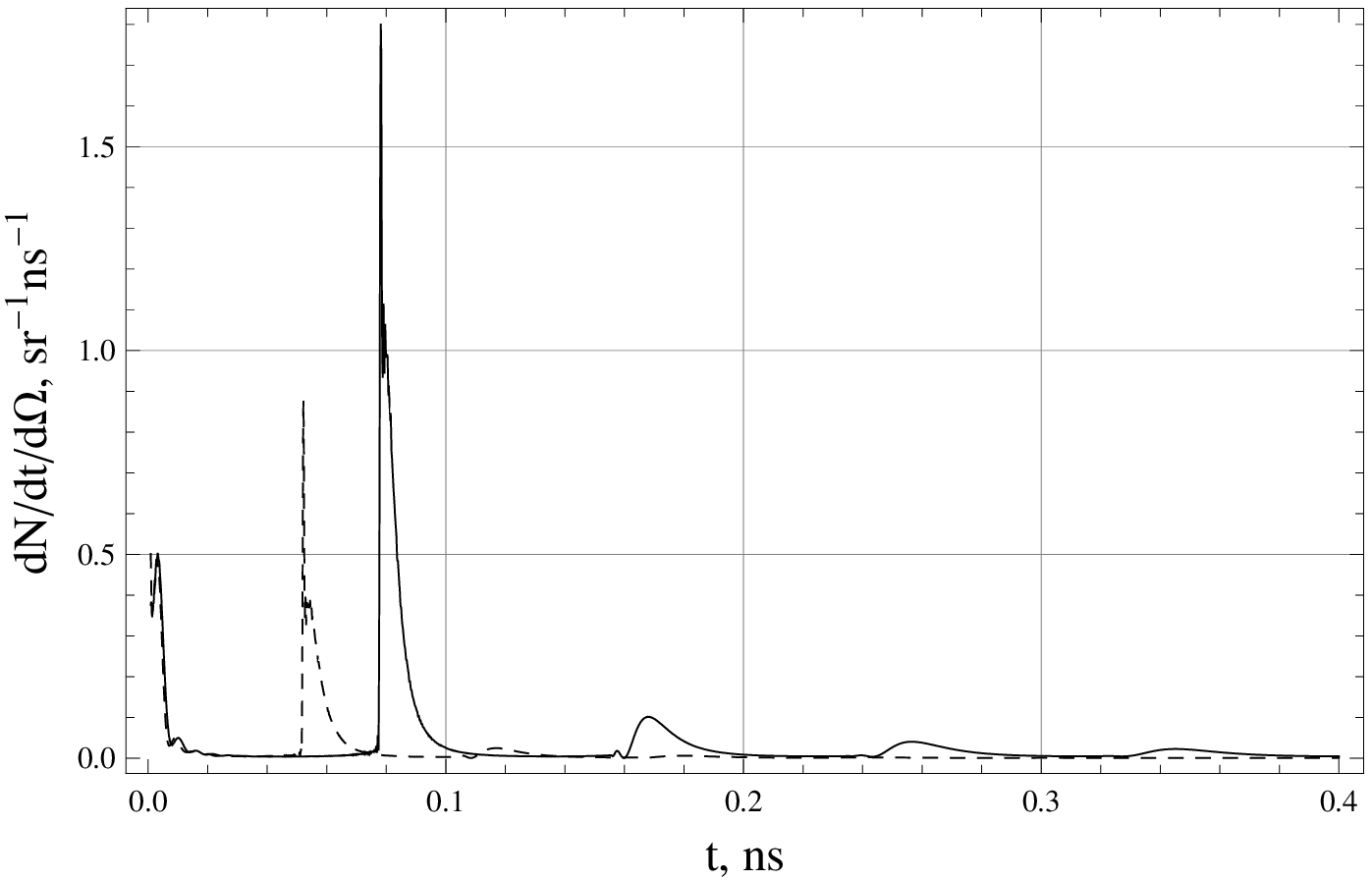}} \resizebox{80mm}{!}{\includegraphics{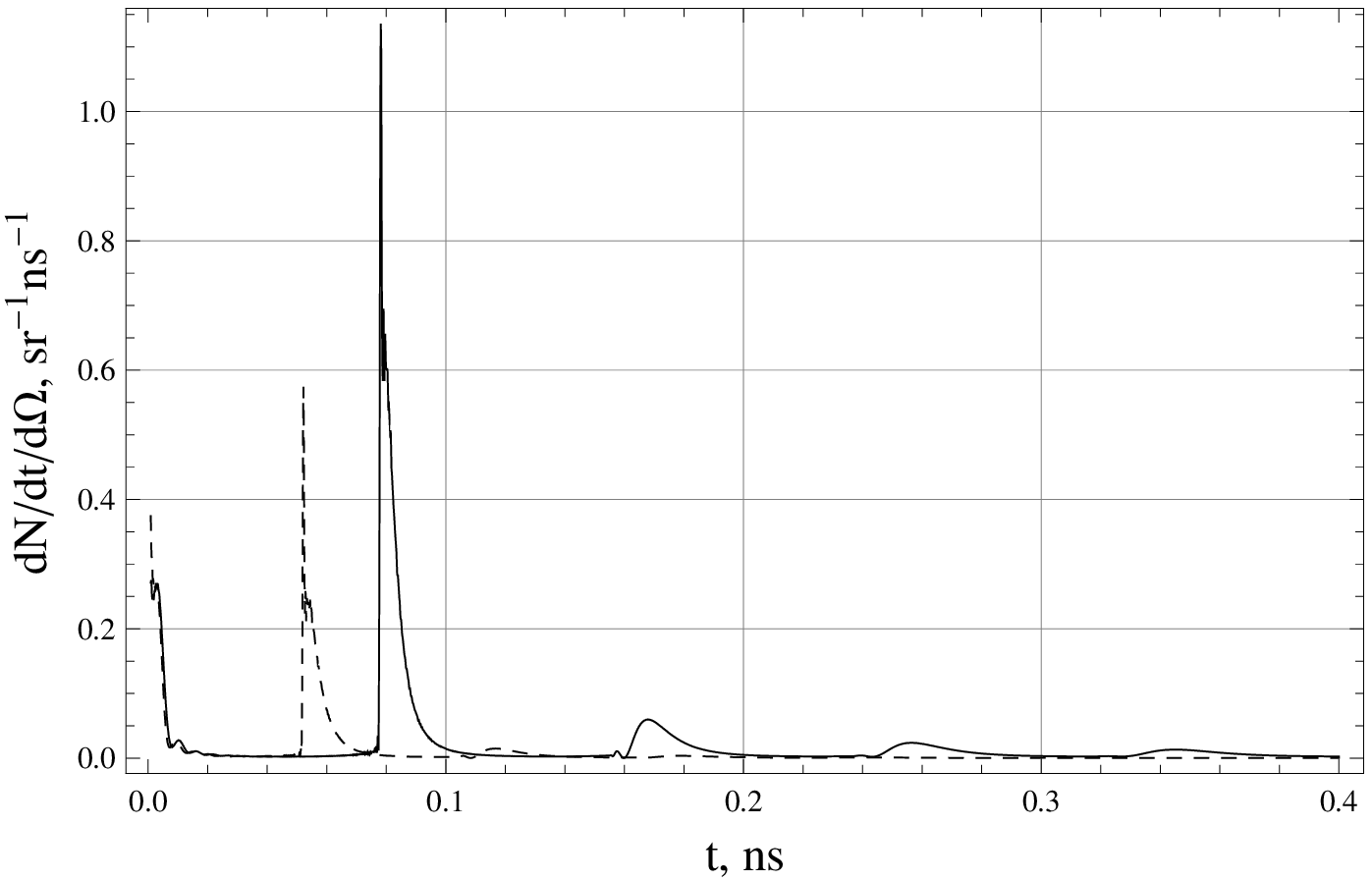}}\\
\end{tabular}

\small{Fig. 4. The PXR in diffracted direction in the Bragg geometry in LiF crystal for electron energies 10 GeV (right), 1 GeV (left): $\beta_1=-45$ (solid line), $\beta_1=-30$ (dashed line) (diffraction plate (5,5,5)).}

 \end{center}
\end{figure}

Now let us consider the PXR in Bragg case (fig. 3). In this case, side by side with electro-magnetic wave emitted in the forward direction, an electromagnetic wave emitted by a charged particle in the diffracted direction and leaving the crystal through the surface of particle entrance can be observed. By matching the solutions of Maxwell's equations on the crystal surfaces  we can get the formulas for the amplitude in Bragg diffraction case for forward directions
\begin{eqnarray}
\label{1.6}
A_{\vec{k}}^s(\omega)&=&\frac{Q\omega}{c^2}(\vec{e}^s \vec{v})\sum_{\mu=1,2}\gamma^0_{\mu s}e^{i\frac{\omega}{\gamma_0}\varepsilon_{\mu s}L}\nonumber\\
&\times&\left[\frac{1}{\omega-\vec{k}\vec{v}}-\frac{1}{\omega-(\vec{k}+\kappa_{\mu s}\vec{N})\vec{v}}\right]\left[e^{i(\omega-(\vec{k}+\kappa_{\mu s}\vec{N})\vec{u})\frac{L}{c\gamma_0}} -1\right]
\end{eqnarray}
and diffracted directions
\begin{eqnarray}
A_{\vec{k_\tau}}^s(\omega)&=&\frac{Q\omega}{c^2}(\vec{e}^{s\tau} \vec{v})\sum_{\mu=1,2}\gamma^\tau_{\mu s}\nonumber\\
&\times&\left[\frac{1}{\omega-\vec{k_\tau}\vec{v}}-\frac{1}{\omega-(\vec{k_\tau}+\kappa_{\mu s}\vec{N})\vec{v}}\right]\left[e^{i(\omega-(\vec{k_\tau}+\kappa_{\mu s}\vec{N})\vec{u})\frac{L}{c\gamma_0}} -1\right],
\end{eqnarray}
where
\begin{equation}
\label{1.9}
\begin{split}
&\gamma^0_{1(2)s}=\frac{2\epsilon _{2(1)s}-\chi_0}{2\epsilon _{2(1)s}-\chi_0-(2\epsilon
_{1(2)s}-\chi_0)\exp\big(\frac{i\omega}{c\gamma_0}(\epsilon_{1(2)s}-\epsilon_{2(1)s})L\big)},\\
&\gamma^{\vec\tau}_{1(2)s}=\frac{-\beta_1C_s\chi_{\vec\tau}}{2\epsilon _{2(1)s}-\chi_0-(2\epsilon
_{1(2)s}-\chi_0)\exp\big(\frac{i\omega}{c\gamma_0}(\epsilon_{1(2)s}-\epsilon_{2(1)s})L\big)}.\\
\end{split}
\end{equation}
With the help of expressions \eqref{1.8} --- \eqref{1.9} time properties of the PXR can be investigated.

In the subsequent analysis it should be taken into account the parameter $|\alpha_B|\sim|\omega-\omega_B|$ rising the PXR intensity decreases rapidly, i. e. the radiation is concentrated near Bragg frequency $\omega_B$. This gives us an opportunity to represent the rate of photon emission in the following form
\begin{equation}
\frac{\partial^2 N}{\partial t\partial\Omega}=\frac{I(\vec r,t)}{\hbar\omega_B}.
\end{equation}

The time evolution of the PXR in diffracted direction is introduced on fig.~4.
Due to the small values of the group velocity in the antisymmetric diffraction  scheme the time of emission lies in the nanosecond range.
Moreover the radiation pulse has complicated structure: it undergoes damping oscillations. The period and the height of radiation peaks increase with $\beta_1$.

It should be noted according to \cite{VG1989,VG1999,VGAA2011} time duration of the quasi cherenkov radiation produced by relativistic particles in optical and microwave ranges in the case $\chi_0,\chi_\tau\ll1$ describe by the formulas analogous to the above.

It should be mentioned that the above discussed phenomena can not
be described in the framework of the PXR theory using the first
Born approximation (though there were a lot of attempts to apply
it for consideration of the PXR spectral-angular properties).

For observation of oscillations, one should either register the moment of particle entrance into the crystal or use a short bunch of
  particles with duration much shorter than the oscillation period. In the X-ray range, such situation is
  typical of electron bunches, which are applied for creating X-ray FELs (FLASH \cite{flash}, XFEL \cite{xfel}, LCLS \cite{lcls}, SPring-8 \cite{spring8}, SwissFEL \cite{swissfel}). (The bunch duration in such
  FELs is tens-hundreds of femtoseconds). If the bunch duration is large in comparison with the duration
  of the radiation pulse or the time of the electron entrance into the crystal is not registered, which occurs in a conventional
  experimental arrangement, then the intensity $I(t)$  should be integrated over
  longer  observation time intervals. As a result, we, in fact, obtain the expression
  integrated over all frequencies, i.e., an ordinary stationary angular distribution of radiation.
  If the response time  of the devices detecting $t_D$ (or  the flight time of the particle in a crystal $t_p$,
  or the bunch duration) is comparable with the oscillation period, then $I(t)$ should be integrated over the interval
  $t_D$. In this case oscillations will disappear, but we will observe the decrease in the intensity of radiation from the crystal.

\section*{Conclusions}
The formulas which describe the time evolution of radiation
produced by a relativistic particle moving in a crystal are
derived. It is shown that the conditions are realizable under
which the parametric X-ray radiation lasts much
longer than the time $t_p$ of  the particle flight through the
crystal. It is shown that such long duration of the radiation makes possible the detailed experimental investigation of the complicated time structure of the parametric X-ray pulses, generated by femtosecond electron bunches, which are available with modern acceleration facilities.

\end{document}